\documentclass{article}

\usepackage[final, nonatbib]{nips_2018}

\usepackage[utf8]{inputenc} 
\usepackage[T1]{fontenc}    
\usepackage{hyperref}       
\usepackage{url}            
\usepackage{booktabs}       
\usepackage{amsfonts}       
\usepackage{nicefrac}       
\usepackage{microtype}      
\usepackage[export]{adjustbox}

\title{Glottal Closure Instants Detection From Pathological Acoustic Speech Signal Using Deep Learning}

\author{
  Gurunath Reddy M, Tanumay Mandal, Krothapalli Sreenivasa Rao \\
  Indian Institute of Technology Kharagpur, India\\
  \texttt{mgurunathreddy@sit.iitkgp.ernet.in} \\
}

\begin{document}

\maketitle

\begin{abstract}
In this paper, we propose a classification based glottal closure instants (GCI) detection from pathological acoustic speech signal, which finds many applications in vocal disorder analysis. Till date, GCI for pathological disorder is extracted from laryngeal (glottal source) signal recorded from Electroglottograph, a dedicated device designed to measure the vocal folds vibration around the larynx. We have created a pathological dataset which consists of simultaneous recordings of glottal source and acoustic speech signal of six different disorders from vocal disordered patients. The GCI locations are manually annotated for disorder analysis and supervised learning. We have proposed convolutional neural network based GCI detection method by fusing deep acoustic speech and linear prediction residual features for robust GCI detection. The experimental results showed that the proposed method is significantly better than the state-of-the-art GCI detection methods.               
\end{abstract}

\section{Introduction}
\label{intrd}

Glottal closures are the instants of significant excitation to the vocal tract system occurs during closure of vocal folds for every cycle of vocal fold vibration during phonation. For healthy vocal folds, the glottal closures can be approximated with a sequence of strong impulses~\cite{rabiner1978digital} due to abrupt closure of vocal folds, where as a disorder vocal fold produces a weak or smeared excitation due to incomplete closure of vocal folds~\cite{yamauchi2016quantification}. Hence, the methods developed to detect GCI from speech signal recorded from healthy vocal folds cannot be directly applied on the pathological speech. Therefore, we can find significantly less attempts to directly extract the GCI from pathological speech. An alternative to the acoustic speech signal is the laryngeal (Electroglottograph or EGG) signal, which is a low frequency signal recorded by measuring the impedance across the larynx by passing a weak electrical current with the help of two electrodes~\cite{childers1985critical}. The EGG signal thus captures the vocal folds activity free from vocal tract resonances which is prominent in raw speech. The negative peaks in the derivative of EGG co-insides with the GCI hence, we can find significantly many works based on EGG for detecting GCI from healthy and disordered vocal folds~\cite{deshpande2018effective, awan2013effect, thomas2009sigma, rothenberg1988monitoring, mittal2014study, mecke2012comparing, chen2013speech}. It is not always possible to record the EGG from the larynx, requires a dedicated device and requires at least basic skills to record EGG from the device. Where as acoustic speech can be easily recorded even with a microphone in a hand held device such as a smart phone, which can be processed in device to give the preliminarily assessment before consulting a doctor or the processed data can be sent to a pathologist for further analysis. Hence, there is a great need for GCI detection directly from speech signal. 

\textbf{Studies based on Modal and Pathological Speech.} We can find many algorithms for GCI detection from modal or normal speech. Most of the available GCI detection methods relies on a representative signal from speech which emphasizes the locations of glottal closure instants. Further, the GCI are detected from the peaks of representative signal either by eliminating spurious instants or by picking genuine GCI using hand crafted heuristics. Most of the methods prefer linear prediction residual (LPR) as a representative signal~\cite{naylor2007estimation, thomas2012estimation, prathosh2016cumulative, koutrouvelis2016fast}, which is a correlate of glottal source signal based on source filter model of speech production~\cite{fant2012acoustic}. Other methods exploit the properties of impulsive nature of excitation~\cite{murty2008epoch, drugman2009glottal, d2011glottal, khanagha2014detection} to derive the representative signal. Recently, \textit{Rachel et.al}~\cite{rachel2017estimation} and \textit{Kebin Wu et. al}~\cite{wu2017gmat} showed their GCI detection method tuned for modal speech as an application to pathological speech on a limited data, which consists of one type of disorder: dysphonia. It should be noted that afore mentioned methods are developed for modal/normal speech and heavily depends on the choice of representation signal, model assumptions to extract the GCI. Also, they depend heavily on signal processing pipe lines requires manual tunning of parameters and hand crafted heuristics specific to dataset to reliably detect the GCI from speech. Recently, classification based data driven model is proposed for GCI detection from modal speech~\cite{matouvsek2017classification, matouvsek2018glottal}. It should be noted that the model is trained with hand crafted features extracted from speech signal and voting classifiers to detect the GCI. In this paper, we propose a deep CNN model, which requires no manual parameter tuning, trained with both raw acoustic and LPR to predict the GCI from pathological speech.   

\section{Vocal Disorder Speech and EGG Dataset}
\label{dataset}

\begin{figure}[h]
        \centering
		\resizebox{14cm}{5cm}{\includegraphics{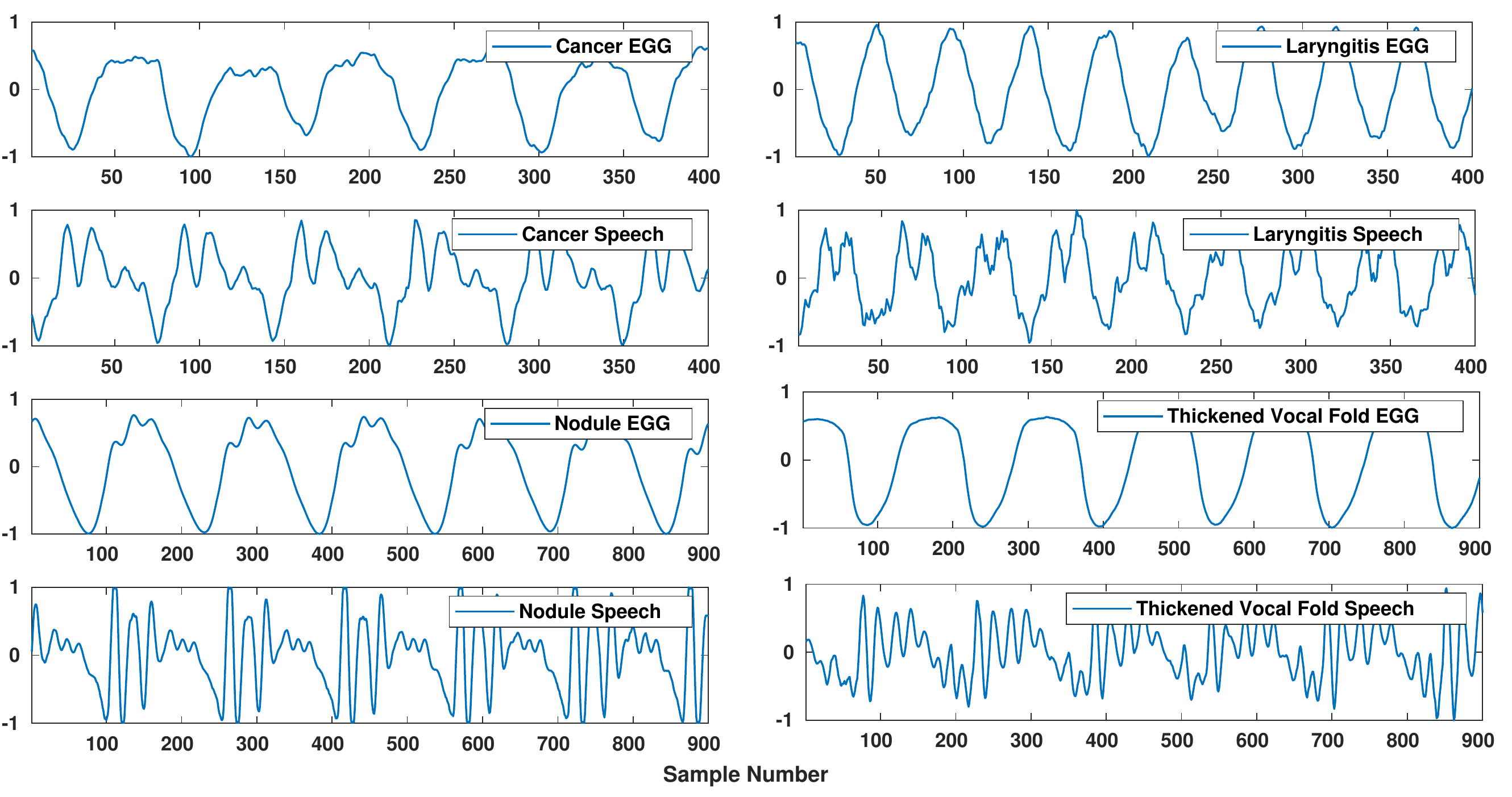}}  
		\vspace{-0.5cm}
        \caption{Pathological speech and the corresponding EGG for various vocal disorders.}
        \vspace{-0.4cm}
        \label{fig:speec_egg}
\end{figure}

For carrying out this study, we collected simultaneous recordings of both EGG and speech signals from the patients who had pathological disorder in vocal folds from B. C. Roy Technology Hospital, Indian Institute of Technology Kharagpur, India. We have collected data from 78 patients registered for diagnosis and treatment from the same hospital. There were 45 male and 33 female patients with the age group of 22-72 years and 12-63 years respectively. The patients suffering from vocal fold disorders are categorized into six types namely vocal nodule (\textbf{N}), vocal polyp (\textbf{P}), laryngitis (\textbf{L}), thickened vocal folds (\textbf{T}), cancer (\textbf{C}) and paralysis (\textbf{PV}) vocal folds. The speech samples are captured using a high quality microphone. The EGG signals are recorded using clinical grade Electroglottograph (procured from TechCadenza, India) device. The acoustic speech and EGG of the sustained vowels /a/, /e/, /o/ are recorded each of them for three times from the participating patients. Sample speech and EGG signals for cancer (\textbf{C}), nodule (\textbf{N}), laryngitis (\textbf{L}) and  thickened vocal folds (\textbf{T}) are shown in Fig~\ref{fig:speec_egg} (due to page constraints, signals are shown for four disorders). 

\section{Proposed GCI detection Method}
\label{prop_method}

\begin{figure}[h]
        \centering
		\resizebox{14cm}{5.5cm}{\includegraphics{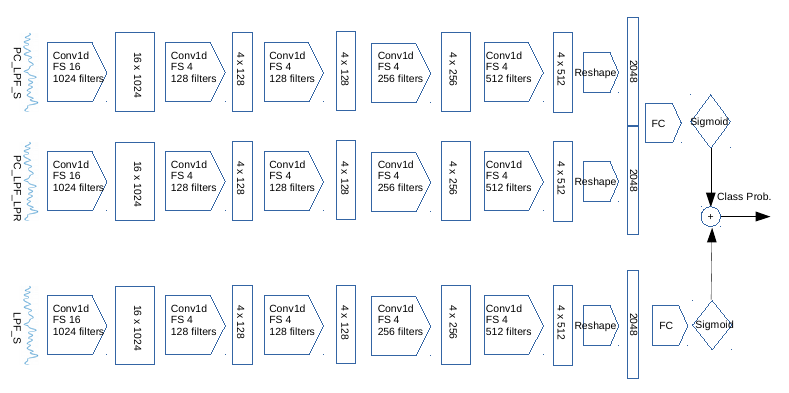}}  
		\vspace{-0.5cm}
        \caption{Multi-column CNN based GCI detection (FS = Filter size, FC = fully connected).}
        \vspace{-0.4cm}
        \label{fig:cnn_model}
\end{figure}

We propose a classification based GCI detection for pathological speech data by training multi-column (parallel) deep CNN models. The proposed CNN based GCI detection architecture is shown in Fig.~\ref{fig:cnn_model}. 

\subsection{Feature Representation, Training and Testing Dataset}
\label{feat_rep}
\begin{figure}[h]
        \centering
		\resizebox{14cm}{5.5cm}{\includegraphics{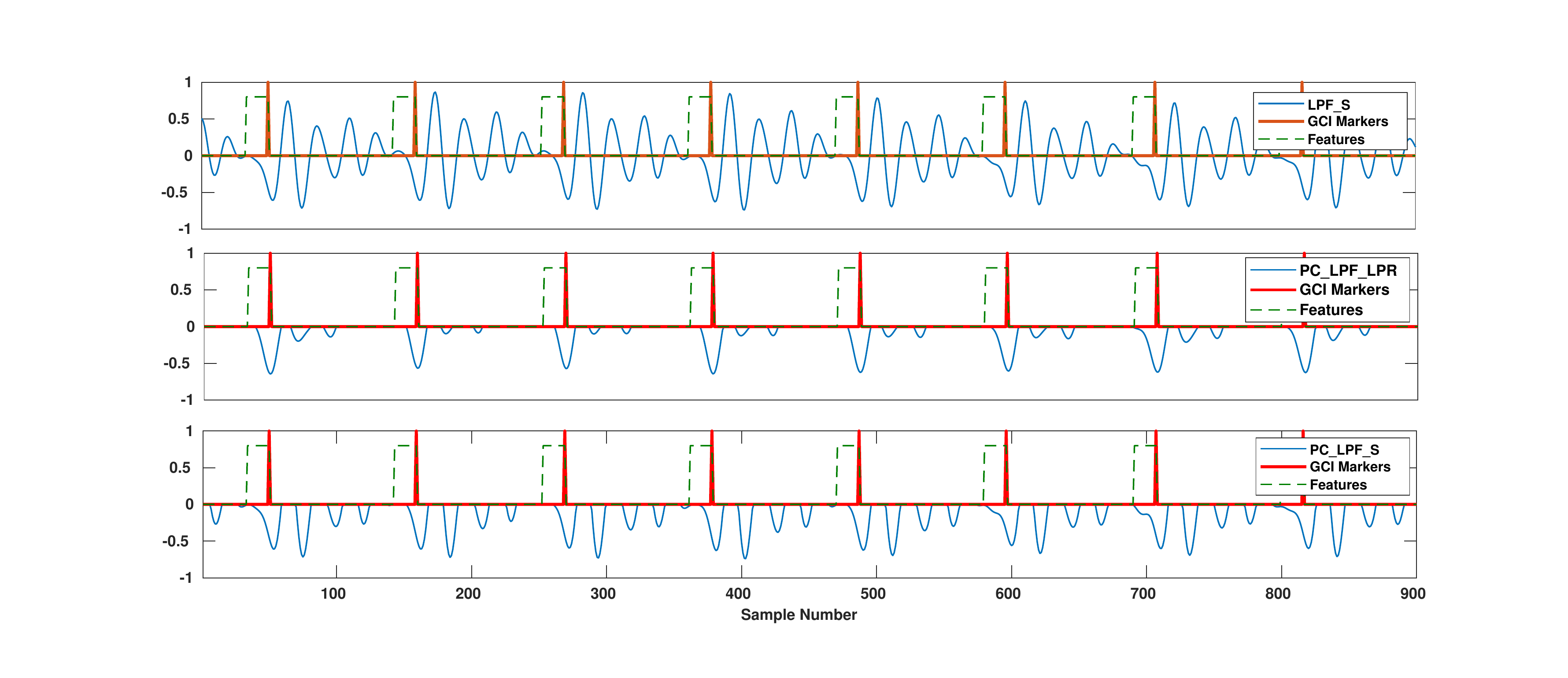}}  
		\vspace{-0.5cm}
        \caption{Input representation, ground truth GCI markers and GCI frame (samples around the GCI).}
        \vspace{-0.4cm}
        \label{fig:feature_rep}
\end{figure}

To leverage the advantage of both raw speech and Linear prediction residual (LPR), we trained the deep models on the following input representations. 1) The low-pass filtered pathological speech signal (\textbf{LPF\_S}). Speech is low-passed since high frequency components do not contribute to the GCI. 2) The low-pass filtered LPR (\textbf{LPF\_LPR}) with LP order 12. The LPR is a noisy residual signal contains unwanted high frequency content hence low-pass filtered. 3) The positive clipped low-pass filtered speech (\textbf{PC\_LPF\_S}) and LPR (\textbf{PC\_LPF\_LPR}). Positive clipped since most of the GCI information is present in the negative portion of the signal~\cite{rengaswamy2016robust}. Low-pass filtering is performed with zero phase, order six low-pass Butterworth filter with cut-of-frequency of 1000Hz. The contemporaneous EGG signal recorded along with the speech is used as reference to manually mark the GCI locations on the speech signal. Speech signals are downsampled to 16 KHz and switched to negative polarity before labeling. The negative peaks in the derivative of the EGG signal is taken as reference to place the GCI markers after compensating the delay between the EGG and corresponding Speech signal. The input representation signals are chunked into frames of 16 samples and assigned a label for presence or absence of GCI. The frame of 16 samples around the GCI (captures GCI slope, shape and amplitude) shown in Fig.~\ref{fig:feature_rep} as dotted box plot is labeled as GCI frame and rest as non-GCI frames for training the model. The training and testing data consist of 62 and 18 pathological speech samples respectively. It should be noted that each sample in the dataset is from unique patient and more than one patient can have same type of vocal disorder.         

\subsection{Multi-Column CNN Model}

The proposed multicolumn deep CNN model for GCI detection is shown in Fig.~\ref{fig:cnn_model}. The final model consists of three CNN networks. Each network is trained with a input representation discussed in~\ref{feat_rep}. Further, each CNN network consists of five convolution layers, each convolution layer is followed by a batch normalization. In our model, we have skipped max pooling layers because we want the model to capture the variations of GCI regions due to stochastic nature of input signal. The resulting feature vector from the CNN layer is connected densely to a sigmoid activation function to predict the posterior classification probability for each frame. The binary cross entropy loss function is optimized by ADAM optimizer with learning rate of 0.0001. The model is trained for 30 epochs until there is no change in validation loss. The weights are initialized from He normal distribution~\cite{kubitschek1962normal}.          

\subsection{Experiments}

\begin{figure}[h]
        \centering
        \label{tab:results}
		\resizebox{12cm}{3.5cm}{\includegraphics{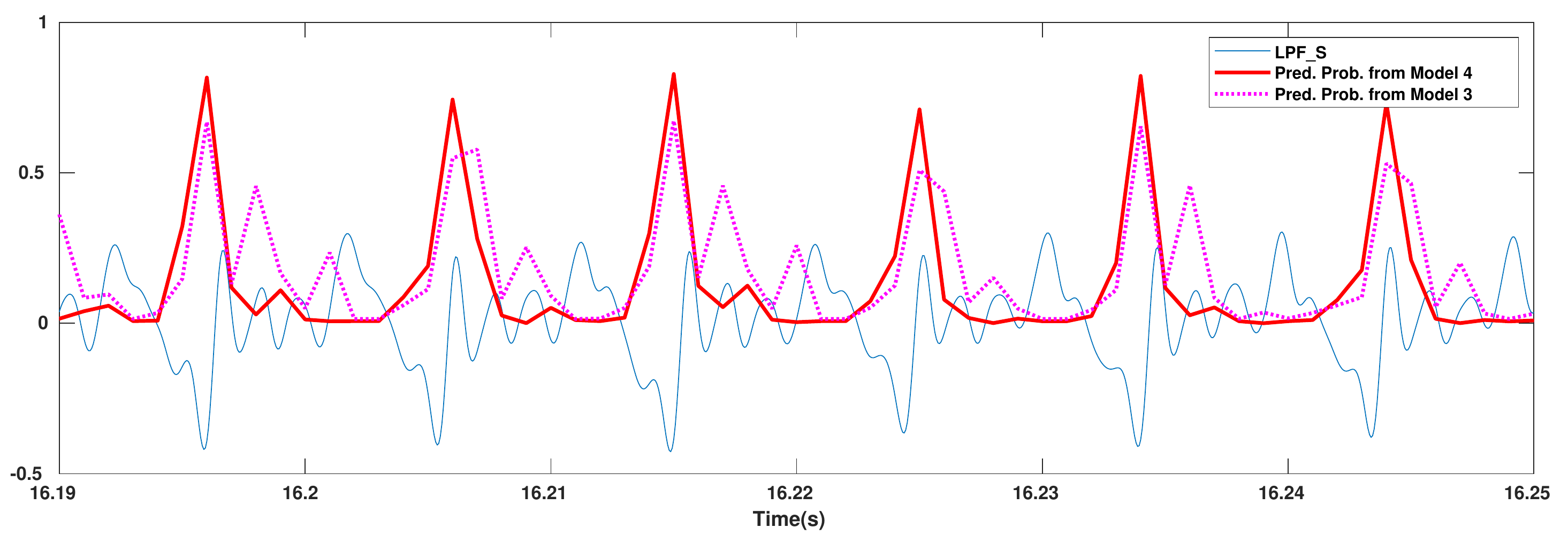}}  
		\vspace{-0.2cm}
        \caption{Posterior class probabilities predicted from \textbf{Model 3} and \textbf{Model 4}.}
        \vspace{-0.4cm}
        \label{fig:post_prob}
\end{figure}

Initially, we trained single column CNN model for each input feature representation discussed in~\ref{feat_rep} separately to evaluate its significance for pathological GCI detection. We denote the models trained with \textbf{LPF\_S}, \textbf{LPF\_LPR}, \textbf{PC\_LPF\_S} and \textbf{PC\_LPF\_LPR} as \textbf{Model 1}, \textbf{Model 2}, \textbf{Model 3} and \textbf{Model 4} respectively. The F1-scores of \textbf{Model 1}, \textbf{Model 2}, \textbf{Model 3} and \textbf{Model 4} are 86.82, 82.94, 86.02, 85.52 respectively. We can observe that models trained with \textbf{LPF\_S}, \textbf{PC\_LPF\_S}, \textbf{PC\_LPF\_LPR} have better F1-score than \textbf{Model 2} trained with \textbf{LPF\_LPR}. A further investigation into the predicted class probabilities from \textbf{Model 1}, \textbf{Model 3} and \textbf{Model 4} revealed that \textbf{Model 3} assigned a little high probability to the non-GCI frames where secondary excitations are prominent, results in false alarms. \textbf{Model 4} assigned very low probability at the low voiced and transition region frames results in high missed rates. Also, \textbf{Model 4} assigns very low probability for the frames with dominant secondary excitation shown in Fig.~\ref{fig:post_prob}, which is good for reducing false alarms. Hence, we trained a joint acoustic-residual model to reap the benefit of both models shown in Fig.~\ref{fig:cnn_model}. In this joint model, we extract the features trained from \textbf{Model 3} and \textbf{Model 4} and append them to a densely connected sigmoid activation function to predict the class probability. Since the model trained with features from low pass filtered speech signal \textbf{LPF\_S} also gave good results, we combine the posterior probabilities of joint acoustic-residual model and model trained with \textbf{LPF\_S} i.e, \textbf{Model 1} in a maximum likelihood sense to predict the final class probability for classification. The frames which achieves class probability greater than or equal to 0.1 are classified as GCI frames. The location of maximum negative peak in the classified frame is considered as glottal closure instant.                 

\section{Evaluation, Results and Summary}
The proposed method is assessed with the reliability and accuracy measures given in~\cite{thomas2009sigma}. Identification Rate (\textbf{IDR}): measures the percentage of GCI for which exactly one GCI is detected. Miss Rate (\textbf{MR}): the percentage of GCI for which no GCI is detected. False Alarm (\textbf{FAR}): the percentage of GCI for which more than one GCI is detected. Identification accuracy (\textbf{IDA}): the standard deviation of the timing error between the detected and the reference GCI location (lower IDA is the better). We compared our method with the popular state-of-the-art GCI detection methods: SEDREAMS~\cite{drugman2009glottal}, DYPSA~\cite{naylor2007estimation} and ZFF~\cite{murty2008epoch}. The evaluation results of the proposed method compared with the other state-of-the-art methods shown in Table~\ref{sample-table}. From Table~\ref{sample-table}, we can observe that the proposed method is significantly better than other methods in detecting GCI with high \textbf{IDR}, low miss rate and false alarms compared to other methods. In summary, we proposed a combined joint acoustic-residual classification based GCI detection method for pathological speech. The evaluation results showed that the proposed method performs significantly better than other methods, which gave the hope to future research on acoustic speech for vocal disorder speech GCI detection and classification.

\begin{table}
  \caption{Comparison of proposed GCI (final model) detection method with other methods.}
  \label{sample-table}
  \centering
  \begin{tabular}{lllll}
    \toprule
    Method     & \textbf{IDR}     & \textbf{MR}     & \textbf{FAR}     & \textbf{IDA}(msec)\\
    \midrule
    Proposed   &89.30         &5.97        &4.70         &0.54     \\
    SEDREAMS   &84.98         &9.06        &6.01         &0.42     \\
    DYPSA      &79.33         &13.42        &7.25         &0.69     \\
    ZFF        &80.88         &10.64        &8.48         &0.56     \\

    \bottomrule
  \end{tabular}
\end{table}

\small

\bibliographystyle{unsrt}
\bibliography{mybib}


\end{document}